# DETERMINATION OF OPTIMAL NUMBER OF CLUSTERS IN WIRELESS SENSOR NETWORKS

Ravi Tandon[1]

[1]Department of Computer Science and Engineering,
Indian Institute of Technology Guwahati, India.
`r.tandon@alumni.iitg.ernet.in`

## ABSTRACT

*Prolonged network lifetime, scalability and efficient load balancing are essential for optimal performance of a wireless sensor network. Clustering provides an effective way of extending the lifetime of a sensor network. Clustering is the process that divides sensor networks into smaller localized group (called clusters) of members with a cluster head. Clustering protocols need to elect optimal number of clusters in hierarchically structured wireless sensor networks. Any clustering scheme that elects clusters uniformly (irrespective of the distance from Base Station) incurs excessive energy usage on clusters proximal and distant to Base Station. In single hop networks a gradual increment in the energy depletion rate is observed as the distance from the cluster head increases[17]. This work focuses on the analysis of wasteful energy consumption within a uniform cluster head election model (EPEM) and provides an analytical solution to reduce the overall consumption of energy usage amongst the clusters elected in a wireless sensor network. A circular model of sensor network is considered, where the sensor nodes are deployed around a centrally located Base Station. The sensor network is divided into several concentric rings centred at the Base Station. A model, Unequal Probability Election Model (UEPEM), which elects cluster heads non-uniformly is proposed. The probability of cluster head election depends on the distance from the Base Station. UEPEM reduces the overall energy usage by about 21% over EPEM. The performance of UEPEM improves as the number of rings is increased.*

## 1. INTRODUCTION

Wireless sensor networks are spatially distributed autonomous system of sensor networks that are deployed for environment supervision, health monitoring, military surveillance, etc. They consist of several wireless sensor motes that collect information from their surroundings and route it to a sink, also called Base Station. One of the primary restrictive factors that affect the performance of wireless sensor networks is limited energy of sensor motes. Consideration of lifetime of networks becomes essential for any deployment strategy because a sensor network can remain effective as long as it is alive [3]. Parameters such as *connectivity, coverage* and *node availability* depend upon the lifetime of a network. Recharging the batteries of sensor nodes is not feasible in many cases due to several reasons (too many nodes, hostile environment etc.)

Hierarchical clustering and data aggregation [11] are two similar approaches that are widely used for prolonging the lifetime of sensor networks. Clustering protocols [5, 8, 13, 15, 17–19] divide the sensor network into separate localized groups called clusters. The network consists of two types of sensor nodes viz. cluster heads and member nodes. The member nodes collect data from the environment and send it to the cluster head. Since the data from the same cluster may have high redundancy due to localized information, data compression is performed by the cluster heads. This process is called data aggregation. The aggregated data is then sent to the Base Station. Since cluster heads aggregate and thus reduce the data to be sent to the Base Station the overall energy usage is reduced.

Communication with the Base Station can be performed in two different ways viz. single-hop communication [5, 19] and multi-hop communication [15, 18]. In single-hop communication every sensor node can reach the Base Station directly, also called Direct Transmission Mode. In multi-hop communication sensor networks route the message using specific routing protocols. Minimum Transmission Energy (MTE) protocols use multi-hop for routing data messages to the Base Station. A special case of multi-hop is a two-hop network. The first hop is from member sensor nodes to cluster heads and the second is from the cluster heads to the Base Station. The primary problem with both the categories of protocols is the creation of energy holes [14, 17]. In single-hop protocols the energy usage increases as the distance from the Base Station increases. In multi-hop networks the energy usage increases as the distance to the Base Station decreases. Multi-hop protocols create hot-spots [14, 16] in the vicinity of Base Station.

Wireless sensor networks can be broadly classified into two major categories viz. homogeneous and heterogeneous sensor networks [9]. Sensor nodes in homogeneous sensor networks have similar capabilities in terms of energy, hardware and processing capabilities. Cluster heads are rotated periodically in order to balance energy usage. However, periodic rotation of cluster heads does not solve the problem of uneven distribution of energy usage in cluster heads with respect to distance from Base Station. Heterogeneous sensor networks consist of sensor nodes with different energies, processing and transmission capabilities. Cluster heads have higher initial energies and can communicate over longer distances. However, fault tolerance is a major issue in heterogeneous sensor networks. Failure of cluster heads may lead to disconnection in sensor network. Homogeneous networks elect cluster heads from a large number of eligible cluster heads and have a higher tolerance towards failure of sensor nodes.

Unequal clustering [8,14,16,17] is an extensively used scheme for dealing with non-uniform energy usage amongst sensor nodes. Multi-hop networks form clusters of decreasing sizes as the proximity to the Base Station increases. However, overall latency for transport of information to the Base Station is high in multi-hop networks. Also they rely heavily on a backbone of cluster heads, which may be subject to failures. Two-hop networks do not have to deal with a construction of path to the Base Station and have a low latency while communicating information to the Base Station. Therefore the case of two-hop homogeneous sensor networks is considered and an analytical solution to the problem to unequal energy consumption in sensor nodes with respect to the distance from Base Station is proposed.

Inefficiencies in the election of optimal number of cluster heads in the existing schemes [2, 5, 7, 17] are analyzed. To overcome these issues a model called Unequal Probability Election Model (UEPEM) is used. The rest of the paper is organized as follows. Section 2 is a review of relevant protocols and models that have been developed to elect cluster heads efficiently. Section 3 provides the energy model. Section 4 contains the network model and its theoretical analysis. Section 5 describes the Unequal Probability Election Model. Section 6 presents the theoretical evaluation of our model and its comparison with Equal Probability Election Model (EPEM). Section 7 presents an extension of the model in a heterogeneous environment. Section 8 consists of conclusion of the work and future research prospects.

## 2. RELATED WORK

Several works [2, 7, 17] have analyzed the distribution of uneven energy usage in a two-hop wireless sensor network. EECS [17] elects cluster heads stochastically for role rotation. A cluster head consumes energy either in aggregation and reception of data from their member nodes or in transmission of data to the Base Station. As the distance of a cluster head from the Base Station increases the energy spent for transmission of data to the Base Station increases too. To counterbalance this problem the cluster size is reduced as the distance increases from the Base Station.

The authors of [7] have devised a scheme that finds an optimal number of cluster heads for a particular region depending on the distance from the Base Station. They assume a linear gradient function that elects higher number of cluster heads as the distance from the Base Station increases. The protocol is called Cluster-based self-Organizing Data Aggregation (CODA).

The authors of [2] have used an approach similar to [5] to determine the optimal number of cluster heads. They compute an average of energies required by the cluster heads closest and farthest from the Base Station. Each cluster head then adjusts its probability of becoming a cluster head depending on the energy consumed in the current round. Their protocol, called LEACH-B, has an effect an inverse effect than EECS and CODA. The probability of cluster head election decreases as the distance of a sensor node from the Base Station increases. This results in larger cluster sizes as the distance from the Base Station increases.

Other protocols such as [8, 10, 14, 16] approach the problem of unbalanced energy consumption by considering a multi-hop scheme. EEUC [8] elects clusters of a lower size to prevent the hot-spot problem closer to the Base Station. The authors of [16] suggest a non-uniform node distribution scheme in order to keep the energy usage equal in each of the rings ("coronas", see Fig. 1). The authors of [14] suggest a scheme analytically wherein a larger number of clusters are formed closer to the Base Station. However, the assumption that clusters remain static over the lifetime may lead to inefficiencies in energy consumption.

Other schemes such [1, 5, 19] rely on rotation of cluster heads based on the residual energy of sensor nodes. While rotation of cluster heads provides efficient load-balancing among sensor nodes that get elected as cluster heads, cluster head election still remains inefficient. Protocols such as [15, 18] form a tree like structure and transmit data over multiple levels of hierarchy. Other protocols rely on MTE (Minimum Transmission Energy) [4, 12] for energy efficient routing. While this leads to efficient energy utilization in sensor nodes far away from the Base Station, nodes closer get depleted of energy faster.

## 2.1 Drawback of earlier approaches

A circular network is considered with the Base Station located at the centre. The sensor network is divided into M rings of equal width (section 4). The innermost ring is ring-0. The inner and outer boundary for the $i_{th}$ ring are concentric circles of radii ir and (i+1)r respectively. EECS [17] elects clusters with a probability $T$. The expected number of cluster heads in the $i_{th}$ ring is given by

$$E[N_{CH}]_i = \rho \, Area_i \qquad (1)$$

$$E_{CH} = \frac{NT}{A_{net}} \pi \, r^2 (2i + 1) \qquad (2)$$

Hence, as the distance from the Base Station increases the number of elected cluster heads increase.

$$E[Total]_i \approx l(N_i(2E_{elec} + E_{DA}) + k_i \epsilon_{mp} d_{toBS}^4 + \epsilon_{fs} d_{toCH}^2 N_i) \qquad (3)$$

When $d^4_{toBS}$ becomes very large then the overall energy consumption depends solely on the number of cluster heads in the $i_{th}$ ring. EECS works well when $d^4_{toBS}$ is not as large. It reduces the value of $d^2_{toCH}$, by keeping cluster sizes smaller and thereby reducing the energy consumption in the $i_{th}$ ring (when $i$ is sufficiently large enough). Hence, an optimal number of cluster heads can be determined only by considering the energy usage pattern. When the clusters are formed at high distance from the Base Station then the energy consumed in transmitting data to the Base Station is higher than the energy consumed for aggregating and receiving data. Hence, electing lower cluster heads tends to aggregate a larger amount of information locally and the information required to be sent to the Base Station is reduced. This reduces the overall energy consumption. The energy consumption per cluster head is higher. However, efficient role rotation [5] and election of cluster heads based on residual energy [1, 19] prevents low energy sensor nodes from becoming cluster heads.

## 3. ENERGY MODEL

This study assumes a simple model for the radio hardware where the transmitter dissipates energy for running the radio electronics to transmit and amplify the signals, and the receiver runs the radio electronics for reception of signals [5]. Multipath fading model ($d^4$ *power loss*) for large distance transmissions and the free space model ($d^2$ *power loss*) for proximal transmissions are considered. Thus to transmit an *l*-bit message over a distance *d*, the radio expends:

$$E_{Tx}(l, d) = E_{Tx-elec}(l) + E_{Tx-amp}(l, d) \quad (4)$$

$$E_{Tx-elec}(l) = l\, E_{elec} \quad (5)$$

$$E_{Tx-amp} = l\epsilon_{fs}d^2, when\ d < d_o \quad (6)$$

$$E_{Tx-amp} = l\epsilon_{mp}d^4, when\ d \geq d_o \quad (7)$$

To receive an *l*-bit message the receiver expends:

$$E_{Rx}(l) = lE_{elec} \quad (8)$$

To aggregate *n* data signals of length *l*-bits, the energy consumption was calculated as:

$$E_{DA-expend} = l\, n\, E_{DA} \quad (9)$$

The radio channel is assumed to be symmetric, so the cost of transmitting a signal from A to B is same as that of transmitting a signal from B to A.

**Table 1: Energy Model Parameters**

| Parameter | Value |
| --- | --- |
| Energy for data aggregation ($E_{DA}$) | 5nJ/bit/signal |
| Initial Node Energy | 0.5J |
| Electronic Energy ($E_{elec}$) | 50nJ/bit |
| Amplification energy for free space model ($\epsilon_{fs}$) | 10pJ/bit/$m^2$ |

| Amplification energy for multi path fading model ($\epsilon_{mp}$) | 0.0013 pJ/bit/$m^4$ |
|---|---|
| Threshold distance ($d_o$) | 87m |
| Packet Size ($l$) | 500 Bytes |

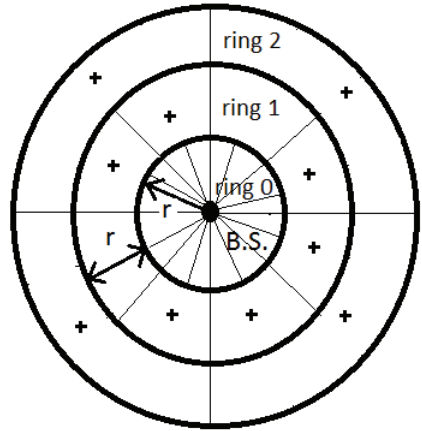

Figure 1. A circular network model that consists of M (M=3) concentric rings around a centrally placed Base Station. The clusters have been approximated to pie shaped regions. '+' denotes a cluster head. Each circular ring has a thickness of r, where $Mr = R_{net}$. Notice that the number of clusters decreases as distance from the Base Station increases. Consequently, cluster size decreases as distance from the Base Station increases.

## 4. NETWORK MODEL

This section describes the network model and other basic assumptions.

1. *N* sensors are uniformly dispersed within a circular field of area $A_{net}$ radius $R_{net} = \frac{A_{net}}{\pi}$. The Base Station is stationed at the centre of the circular region. The number of sensor nodes *N* to be deployed depends specifically on the application.

2. The sensor nodes are considered to be stationary. Each sensor node can communicate with the Base Station directly.

3. Although a Berkley sensor mote has over 100 different power levels [6] continuous power levels for simplicity as in [5] is assumed.

4. Communication is symmetric and a sensor can compute the approximate distance based on the received signal strength if the transmission power is known.

5. All sensors are location-unaware, i.e. not equipped with GPS.

6. All sensors are homogeneous, i.e., they have the same capacities.

All the sensor nodes have a particular identifier (ID) allocated to them. Each cluster head coordinates the MAC and routing of packets within their clusters. Thus the clusters are synchronized by cluster heads perfectly and there is no loss of energy while a sensor node is awake and waiting for its turn. The sensor network is divided into *M* rings of equal thickness, *r*, where $Mr = R$ (see Fig. 1). A data logging application is assumed where sensor nodes sense data and send it to their respective cluster heads. The cluster heads send data to the Base Station.

The rate of data being sensed is considered to be uniform throughout the sensor network. The medium is assumed to be contention free and control messages between the cluster heads and the sensor nodes are not considered, assuming them to be very short and introduce only a very small overhead. Cluster heads aggregate data in a perfect manner. Therefore, cluster heads send only a single data packet to the Base Station per round.

**Table 2: Symbols used in the derivation of UEPEM**

| Symbol | Parameter Represented |
|---|---|
| $A_{net}$ | Area of sensor network |
| $R_{net}$ | Radius of sensor network |
| N | Number of sensor nodes |
| M | Number of rings |
| r | Thickness of a ring |

Every cluster consists of Voronoi shaped regions. Each ring has $k_i$ cluster heads. Furthermore, in order to simplify the theoretical analysis of sensor network, Voronoi shaped clusters are approximated with pie shaped regions (see Fig. 1), as in [14]. Furthermore, the sizes of each cluster in a single ring are assumed to be same.

### 2.1. Theoretical Analysis

Let $Z_i$ be a random variable denoting distance of cluster head from the sensor node in a cluster for $i_{th}$ ring.

**Lemma 1.** The expected value of $Z_i^2$ is $r^2(2i + 1)2\,k_i$.

*Proof.* Each cluster is assumed to be circular in shape. There are $k_i$ clusters in the $i_{th}$ ring. Each cluster in $i_{th}$ ring has the same size. Therefore

$$k_i \,\pi r_{cluster_i}^2 = A_i \quad (10)$$

Here $A_i = \pi(r_i^2 - r_{i-1}^2) = \pi r^2(2i + 1)$. Cluster radius $r_{cluste_i}$ for the $i_{th}$ ring is

$$r_{cluster_i} = r\sqrt{\frac{2i + 1}{k_i}} \quad (11)$$

The expected value of $Z_i^2$ is calculated as follows.

$$E[Z_i^2] = \int_{\theta=0}^{\theta=2\pi} \int_{r=0}^{r=r_{cluster_i}} \rho(r,\theta) r^2 r\, dr\, d\theta \quad (12)$$

If the density of sensor nodes is uniform throughout the area then $\rho$ becomes independent of $r$ and . It is equal to $\frac{1}{\frac{A_i}{k_i}}$ which equals $\frac{k_i}{\pi r_{cluter_i}^2}$. Using Eq. 11,

$$E[Z_i^2] = r^2(2i + 1)2k_i \quad (13)$$

Let $Y_i$ be a random variable denoting the distance of a cluster head (in the $i_{th}$ ring) from the Base Station.

**Lemma 2.** The expected value of $Y_i^4$ is $\frac{(r^4)((i+1)^6 - i^6)}{3(2i+1)}$.

*Proof.* The expected value of $Y_i^4$ is estimated as follows.

$$E[Y_i^4] = \int_{\theta=0}^{\theta=2\pi} \int_{r=ir}^{r=(i+1)r} \rho(r,\theta) r^4 r \, dr \, d\theta \tag{14}$$

The cluster heads are assumed to be uniformly distributed throughout the network. Since the density of sensor nodes is uniform throughout the area $\rho$ becomes independent of $r$ and $\theta$. It is equal to $\frac{1}{\pi r^2 (2i+1)}$. Solving Eq. 14,

$$Y_i^4 = \frac{(r^4)((i+1)^6 - i^6)}{3(2i+1)} \tag{15}$$

**Lemma 3.** The optimal number of cluster heads in the $i_{th}$ ring is $\sqrt{\frac{(3\epsilon_{fs} N (2i+1)^3)}{2\epsilon_{mp} R_{net}^2 ((i+1)^6 - i^6)}}$

*Proof.* The optimal number of cluster heads for the network model is analytically determined using the computation model and communication models (Section 3). There are $N_i$ sensor nodes in the $i_{th}$ ring out of which $k_i$ are cluster heads. Uniform distribution of sensor nodes in each cluster is assumed. The number of sensor nodes in each cluster is $\frac{N_i}{k_i}$. Each cluster head dissipates energy while receiving data from each of the sensor nodes in its cluster, aggregating the data and then transmitting the data to Base Station. A two-hop model is considered. Each sensor node sends collects equal amount of data per round. The energy consumed by a cluster head in a single round is

$$(E_{CH})_i = l((E_{elec} + E_{DA})\left(\frac{N_i}{k_i} - 1\right) + E_{elec} + E_{DA} \epsilon_{mp} (d_{toBS})_i^4 \tag{16}$$

Using Lemma 2 and Eq.16 the expected value of the energy consumed by a cluster head is calculated as follows.

$$E[(E_{CH})_i] = l((E_{elec} + E_{DA})\left(\frac{N_i}{k_i}\right) + \epsilon_{mp} \left(\frac{(r^4)((i+1)^6 - i^6)}{(3(2i+1))}\right) \tag{17}$$

Member sensor nodes sense data from the environment and send data to their cluster head. The energy consumed by a member sensor node per round is calculated as follows.

$$E(SN)_i = l(E_{elec} + \epsilon_{fs}(d_{toCH})_i^2 \tag{18}$$

Using Lemma 1 and Eq. 18, the expected value of the energy consumed by a sensor node is calculated as follows.

$$E[(E_{SN})_i] = l\left(E_{elec} + \epsilon_{fs}\frac{r^2(2i+1)}{2k_i}\right) \quad (19)$$

Using Eq. 17 and Eq. 19, the total energy consumed by a cluster is estimated as follows.

$$E[(E_{cluster})_i] = E[(E_{CH})_i] + \left(\frac{N_i}{k_i} - 1\right)E[(E_{SN})_i] \approx E[(E_{CH})_i] + \frac{N_i}{k_i}E[(E_{SN})_i] \quad (20)$$

which is equal to,

$$E[(E_{cluster})_i] = l((2E_{elec} + E_{DA})\frac{N_i}{k_i} + \frac{\epsilon_{mp}(r^4)((i+1)^6 - i^6)}{3(2i+1)} + \frac{\left(\frac{\epsilon_{fs}r^2(2i+1)}{2k_i}\right)N_i}{k_i}) \quad (21)$$

Using Eq. 21 the total energy consumed in the $i_{th}$ ring is estimated as follows.

$$E[(E_{Total})_i] = k_i E[(E_{Cluster})_i] \quad (22)$$

$$E[(E_{Total})_i] = l\left((2E_{elec} + E_{DA})N_i + k_i\epsilon_{mp}\frac{(r^4)((i+1)^6 - i^6)}{3(2i+1)} + \frac{\epsilon_{fs}r^2(2i+1)}{2k_i}N_i\right) \quad (23)$$

An observation is that the energy consumed by the sensors in each ring is independent of the sensors in the other rings. The number of cluster heads elected is optimal when the total energy consumed by sensors in the $i_{th}$ ring is minimized. To get the optimal value of the number of clusters Eq. 23 is differentiated with $k_i$, and get $k_{opt}$ as

$$(k_{opt})_i = \sqrt{\frac{(3\epsilon_{fs}N(2i+1)^3)}{2\epsilon_{mp}R_{net}^2((i+1)^6 - i^6)}} \quad (24)$$

## 5. UNEQUAL PROBABILITY ELECTION MODEL

A model called *Unequal Probability Election Model (UEPEM)* where in the probability of a cluster head election depends up on the distance of the sensor node from the Base Station is proposed. Since sensor nodes are not equipped with a GPS the Base Station sends a "hello" message for initiation. Each sensor node then approximates its distance (and thereby to which ring it belongs to) from the Base Station based up on the received signal strength. The probability of election of a cluster head is given by the following Eq. 25,

$$P(CH)_i = \frac{k_i}{N_i} \quad (25)$$

which is equal to,

$$P(CH)_i = \frac{M^2}{R_{net}}\sqrt{\frac{3\epsilon_{fs}(2i+1)}{2\epsilon_{mp}((i+1)^6 - (i)^6)N}} \quad (26)$$

Clusters are formed in the form of Voronoi tessellations similar to those in [4], where each sensor node joins the closest cluster head.

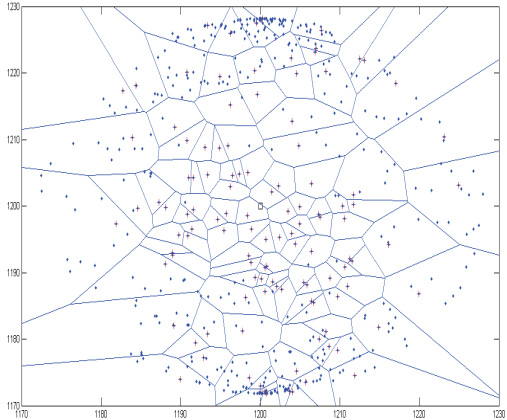 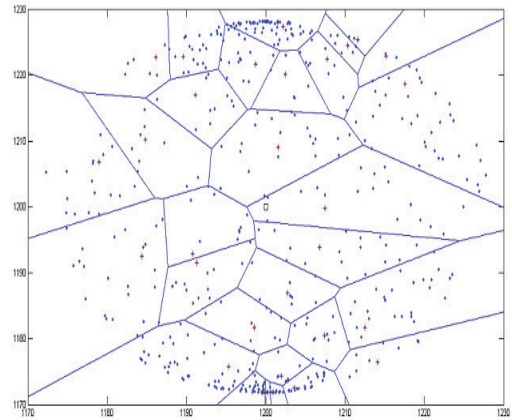

Figure 2: Voronoi tessellations graphs for Unequal Probability Election Model

Figure 3: Voronoi tessellations graphs for Equal Probability Election Model

## 6. THEORETICAL EVALUATION

In this section evaluation of the UEPEM model has been done. For comparison a model that elects cluster heads with a uniform probability $\rho$ (=0.05) is assumed. This model is called *Equal Probability Election Model (EPEM)* as in [10]. Both the models are compared to show that UEPEM succeeds in reducing the energy consumption in every ring around the Base Station. The energy consumed by a cluster head and a sensor node in the $i_{th}$ ($i \geq 1$) around the Base Station is measured. The average cluster size, both in terms of radius and number of nodes within the cluster, is also analytically determined. The number of nodes (*N*) considered are 500. The area of the network ($A_{Net}$) is $25 * 10^4 m^2$. The number of rings is *10*.

### 6.1. Energy Usage

In this subsection the total energy usage by sensor nodes per round for each ring is analyzed. UEPEM minimizes the energy usage on an average by about 21% (see Fig. 4) over EPEM. Interestingly the improvement in the energy usage in the rings closest to the Base Station and furthest away from the Base Station is the largest. Fig.5 shows that for first ring the energy consumed by UEPEM is lesser than half of the energy consumed by EPEM. For the last two rings the energy consumption is reduced by 34% and 46% respectively. This shows that a clustering approach that elects cluster heads uniformly performs poorly in the rings that are either very close or very far away from the Base Station. The reason for higher energy consumption is that EPEM elects cluster heads with a uniform probability, which is suited for the average case (the middle rings). For middle rings (rings 3-6) the average energy spent per sensor node in UEPEM is 0.96 that of EPEM. The rings that are closer to the Base Station incur loss when data is aggregated and received by the cluster heads. Sensor nodes closer to the Base Station consume lesser energy when transmitting data directly to the Base Station. Losses due to direct transmission are higher for rings that are farther away. Another reason for improvement in the total energy consumed by sensor nodes in inner rings in UEPEM is the lower energy consumption by cluster heads (see Fig. 6). The energy consumption in the sensor nodes that are in outer rings in UEPEM improves due to the lesser number of cluster heads (see Fig. 13).

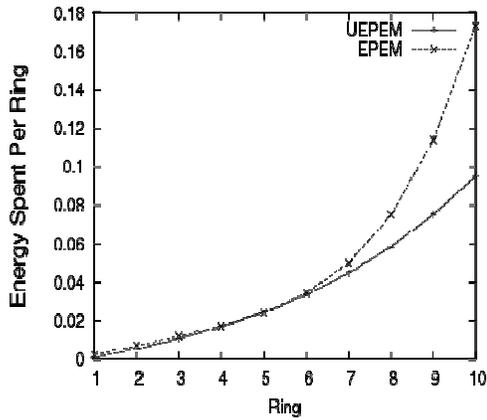
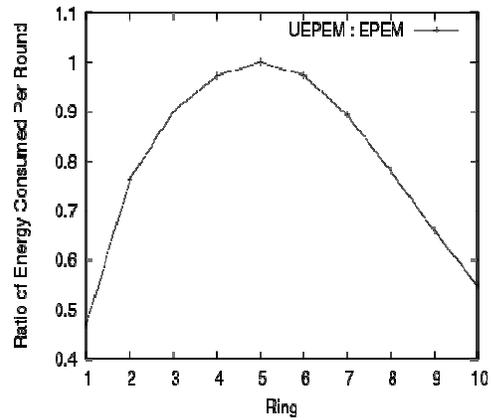

Figure 4. Shows the total energy usage for EPEM and UEPEM for all the sensor nodes in each ring.

Figure 5. Shows the ratio of energy usage for each of the rings around the Base Station.

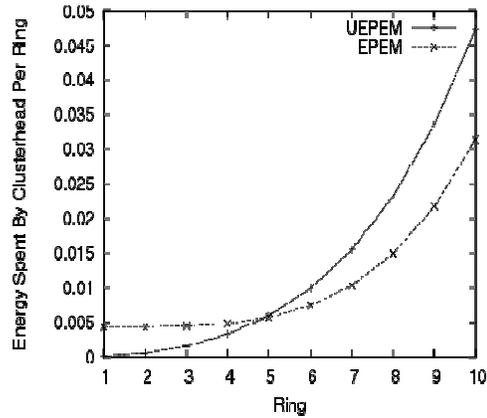
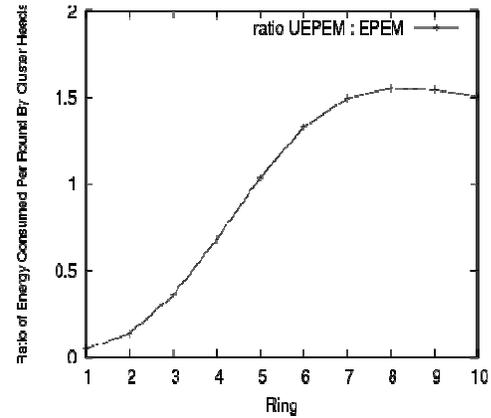

Figure 6. Shows the energy consumed by a single cluster head for both UEPEM and EPEM around the Base Station for each ring.

Figure 7. Shows the ratio of energy consumed by a cluster head for each ring for UEPEM and EPEM.

### 6.2. Energy Consumption By Cluster Heads

In this subsection the energy consumed by the cluster heads is analyzed. The cluster heads in the rings closer to the Base Station consume lesser energy than for UEPEM than those for EPEM refer Fig. 6 and Fig. 7. For the inner five rings (rings 1-5) the energy consumed by a cluster head in UEPEM is 0.45 times that of EPEM. The reason primarily is lower cluster sizes for the rings that are closer to the Base Station. This results in lesser energy consumption for aggregation and reception of energy. For ring 5 (almost the centre) the energy consumption becomes nearly equal, reinforcing our explanation that the optimal cluster head election by EPEM is for the rings with average distance to the Base Station. For the rings that are farther away from the Base Station, the average energy usage is higher for UEPEM. For the outer rings (rings 6-10) the average energy consumed by a cluster head in UEPEM is 48% higher than that for EPEM. However, the overall energy usages by all the sensor nodes in rings that are farther away remain lower (see Fig. 7). This is because the number of cluster heads that are elected are

244

lower (see Fig. 13) for UEPEM. Hence, lesser amount of data is transmitted to the Base Station and more data is locally received and aggregated into a single transmission packet. Such a scheme would be very suitable in applications where either the required accuracy in data is low or the data has very high redundancy.

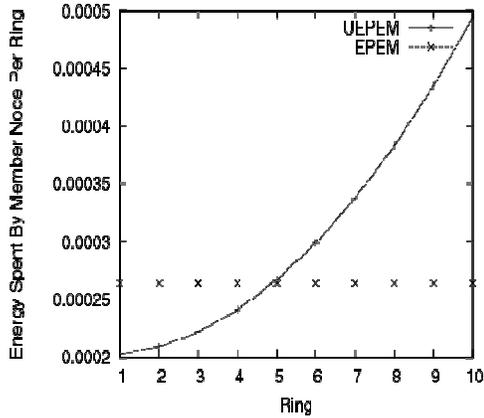 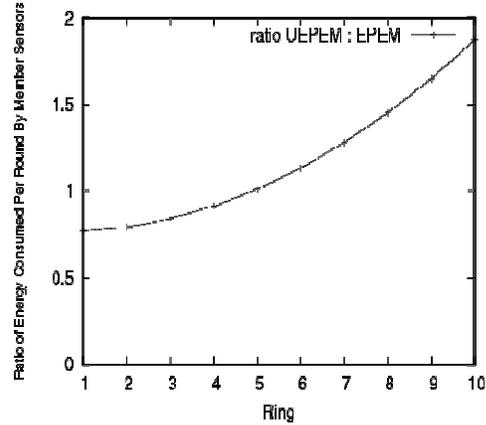

Figure 8. Shows the energy consumed by a single member node for both UEPEM and EPEM around the Base Station for each ring.

Figure 9. Shows the ratio of energy consumed by a single member node for each ring for UEPEM and EPEM.

### 6.3. Energy Consumption By Sensor Nodes

In this subsection the energy consumption of sensor nodes (that are not cluster heads) for different rings are analyzed. They are referred to as member sensor nodes. The pattern of energy consumption is similar to that of cluster head consumption (see Fig. 6 and Fig 7). The energy usage for member sensor nodes that are closer to the Base Station is lower for UEPEM than that for EPEM primarily because the cluster size (see Fig. 11) for the inner rings is smaller. Hence each member sensor node has to spend lesser amount of energy transmitting data to their respective cluster heads. For the inner rings in UEPEM the average energy consumption per round is 0.86 times that of EPEM. We can observe that energy gain for UEPEM is higher in the case of cluster heads than that for member sensor nodes. The energy consumed by member sensor nodes further away from the Base Station is higher for UEPEM. In UEPEM, the member sensor nodes (in the outer rings) on an average consume 1.47 times the energy consumed by member sensor nodes in EPEM. This is because of the high cluster size in the rings that are farther away from the Base Station (see Fig. 11). The overall energy usage is still better (See Fig. 6) because the additional energy consumed by sensor nodes is counter balanced by the energy savings from cluster heads (see Fig. 6).

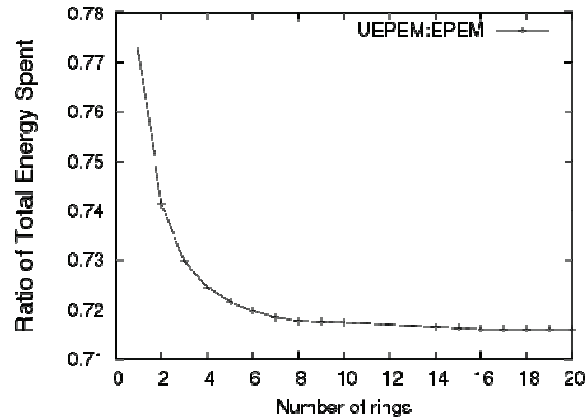

Figure 10.Comparison of ratio of total energy consumed for UEPEM and EPEM for different number of rings around the Base Station.

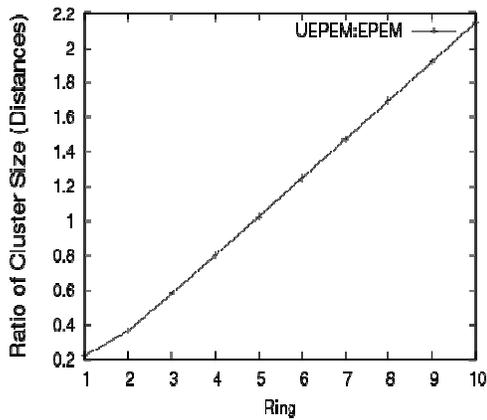

Figure 11.Shows the ratio of cluster radii for UEPEM and EPEM in rings around the Base Station.

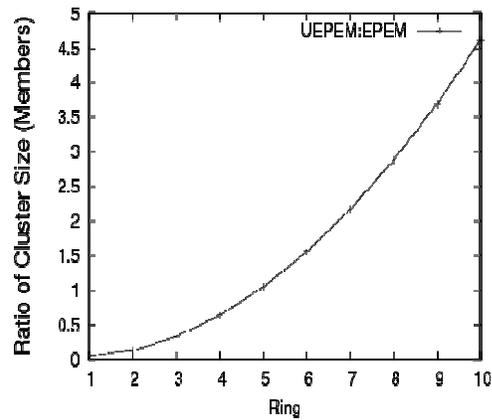

Figure 12.Shows the ratio of cluster size for UEPEM and EPEM in rings around the Base Station.

### 6.4. Total Energy Consumed With Variation of Number of Rings

The number of rings that the sensor network has to be broken into is a design parameter of UEPEM model. In order to determine an accurate value the number of rings (M) is varied from 1 to 20. As the network is broken up into larger number of rings the energy consumption reduces (see Fig. 4). The energy consumed by the sensor nodes per ring for EPEM model is 0.5 Joule. For UEPEM, the average energy consumed varies from 0.39 Joule to 0.365 Joule. The thickness of each ring reduces as the number of rings considered in the UEPEM model increase. The calculated value of $k_{opt}$ is an approximation. In Lemma 2 distance of each cluster head from the Base Station for each has been calculated. As the number of rings increases, the deviation in the value of distance from cluster heads to Base Station decreases. Therefore, the value of $k_{opt}$ can be predicted even more accurately. Hence, the overall energy usage per round decreases as the total number of rings in the sensor network increases for the UEPEM model. The EPEM model is independent of the number of rings considered in the sensor network. The energy usage per round decreases as the number of rings increases. With a single ring the

246

average energy consumed in UEPEM is about 0.77 times that of EPEM model. With 20 rings the ratio average energy consumed down to 0.72, showing an improvement of 5%.

## 6.5. Cluster Size

This subsection presents the analysis of the size of clusters formed in each of the rings around the Base Station. Fig. 11 shows the ratio of cluster radius for UEPEM and EPEM in rings around the Base Station. Fig. 12 shows the ratio of cluster size for UEPEM and EPEM in rings around the Base Station. The cluster sizes for rings that are closer to the Base Station are smaller in size for UEPEM as compared to those for EPEM. For the first five rings cluster size for UEPEM is about 40% smaller than that for EPEM. The average cluster radius for EPEM is *40m* while for UEPEM it varies from *8m* to *85m*. For the last five rings the average cluster size was about 71% high for UEPEM as compared to EPEM. In the inner rings the higher number of clusters is formed. The average cluster size for EPEM is 20 (= $\frac{1}{\rho}$, $\rho$ is 0.05).The cluster size for UEPEM varies from *1* to *90* for UEPEM. For the first ring each sensor node is a cluster head. Therefore, cluster sizes are smaller. For the outer rings the cluster sizes becomes almost 7.5 times higher on an average.

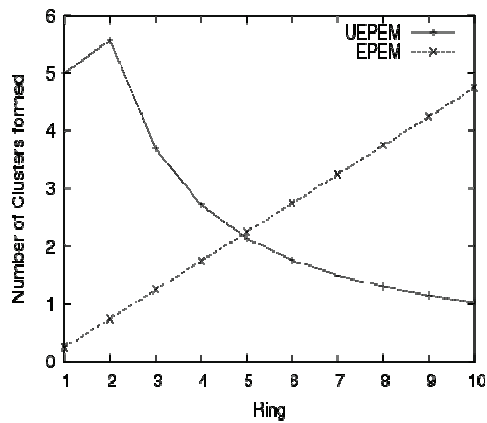
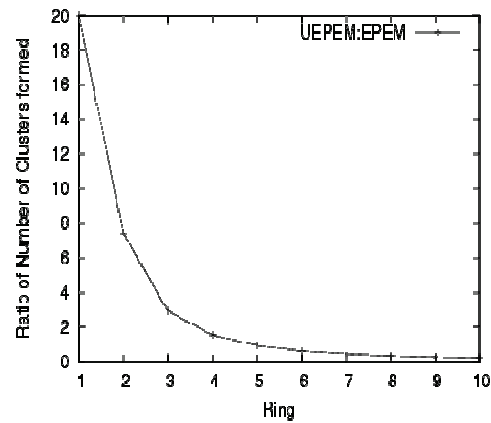

Figure 13.Shows the number of clusters formed as a function of rings around the Base Station.

Figure 14. Shows the ratio of number of clusters formed as a function of rings around the Base Station.

## 6.5. Number of Clusters

This subsection analyzes the number of clusters formed in each of the rings around the Base Station. Fig.13 and Fig. 14 show a comparison of the number of clusters formed for UEPEM and EPEM model. The number of clusters in the inner rings (the first 5 rings) is almost 200% higher for UEPEM. Since the sensor nodes are closer to the Base Station in these rings the cost associated with direct transmission is lesser. Hence more cluster heads get elected. However, as the distance from the Base Station increases the number of clusters decrease for UEPEM. The number of clusters formed for UEPEM (for outer rings 6-10) is about 65% lower than that for EPEM. The number of clusters increases for EPEM as the distance from the Base Station increases. The probability of cluster head election remains constant, while the number of sensor nodes increase as the distance from Base Station increases. Hence more clusters get elected as the distance from the Base Station increases. The ratio of cluster heads elected (UEPEM:EPEM) goes down from 20 to 0.2 (see Fig. 14).

## 7. HETEROGENEOUS SENSOR NETWORKS

The model *Unequal Probability Election Model (UEPEM)* has been designed for a homogeneous sensor network. A heterogeneous sensor network consists of sensor nodes with different processing and transmission capabilities. Two different categories of sensor nodes are assumed. Category I consist of high energy, large distance transmission and data aggregation capacities and category II consist of low energy nodes with short distance transmission capacity. Category II sensor nodes are assumed to be uniformly and randomly deployed in a circular region around the Base Station. The total number of such sensor nodes is dependent on the application. Using UEPEM a scheme for the optimal deployment scheme for category is suggested. The number of category I sensor nodes to deployed in the $i_{th}$ ring is given by Lemma 3. Clusters are formed in a static manner. This leads to a reduction in the overhead during cluster formation.

Thus, a static clustering approach based on the UEPEM can be considered. The deployment of category I sensor nodes can be done according to the value of $k_{opt}$ as suggested by Lemma 3.

## 8. CONCLUSION AND FUTURE WORK

This work analyzes the problem of inefficient cluster head election in a wireless sensor network. The analysis shows that election of cluster heads with a uniform probability as in (Heinzelman, MA Chandrakasan, & Balakrishnan, 2002) leads to inefficient energy usage. For this the sensor network has been divided into concentric rings ($M = 10$) around the Base Station. It is also shown that for rings proximal to the Base Station more cluster heads n be elected (almost 200%). For rings that are farther away from the Base Station the lesser cluster heads should be elected (almost 65%).

An analytic model to predict the number of cluster heads as a function of distance from the Base Station has been suggested. The model is called *Unequal Probability Election Model (UEPEM)*. In this work a theoretical comparison with Equal Probability Election Model (EPEM) is done. UEPEM on an average provides 28% reduction in total energy usage over EPEM. The analysis suggests that higher number of cluster heads should be elected closer to the Base Station. As the distance from the Base Station increases the number of clusters formed decreases and consequently the cluster size increases.

Future research study will focus on simulation of UEPEM and comparison with our protocols such as EECS [17], CODA [17] and LEACH-B [2]. Study of lifetime, cluster characteristics of our model in comparison to EECS, CODA and LEACH-B will be done. To ease the analysis several simplifying assumptions have been made. Firstly, Voronoi cells have been approximated as pie shaped regions and clusters have been assumed to be circular in shape for the calculation of average distance to cluster head. These issues will be addressed in future research. Extending the model to a heterogeneous environment will also be looked into.